\documentclass[a4paper,11pt]{article}
\usepackage{pos}
\usepackage{schemata}

\title{Effective Lagrangians and thermal resonances under extreme conditions}

\author*[a]{Andrea Vioque Rodríguez}
\author[a]{Angel Gómez Nicola}
\author[a]{Jacobo Ruiz de Elvira}

\affiliation[a]{Departamento de Física Teórica and IPARCOS, Univ. Complutense, 28040 Madrid, Spain.}

\emailAdd{avioque@ucm.es}
\emailAdd{gomez@ucm.es}
\emailAdd{jacobore@ucm.es}

\abstract{We analyze various problems related to the physics of hadrons under extreme conditions of temperature and chemical potentials. On the one hand, we show that the thermal resonances $f_0(500)$ and $K_0^*(700)$, generated in the framework of Unitarized Chiral Perturbation Theory $\pi\pi$ and $K\pi$ scattering at finite temperature, play an essential role concerning chiral and $U(1)_A$ restoration. On the other hand, a low-energy effective Lagrangian has been constructed within ChPT at non-zero chemical potential, which we discuss here for  an axial chemical potential.}

\FullConference{%
  10th International Conference on Quarks and Nuclear Physics\\
  8-12 July, 2024\\
  Barcelona, Spain
}

\begin{document}
\maketitle

\section{Introduction}

The main goal of this talk is to study the nature of the chiral transition and its connection to the $U_A(1)$ symmetry restoration using effective theories. Chiral symmetry can be restored by external effects such as temperature. At $\mu_B = 0$, lattice simulations have quite firmly established that the chiral transition is a crossover at $T_c\simeq 155$ MeV for $2+1$ flavors and physical quark masses~\cite{Aoki:2009sc,Bazavov:2011nk}, usually identified as the peak position of the scalar susceptibility. The subtracted quark condensate and chiral susceptibility are the key observables to study the chiral transition.

Another important problem, still under investigation, is whether $U_A(1)$ restoration occurs close to the chiral transition. While for $N_f=2+1$ lattice simulations with physical quark masses, the $U_A(1)$ symmetry remains significantly broken at the chiral transition~\cite{Buchoff:2013nra}, $N_f=2$ analyses near the light chiral limit indicate compatibility between chiral and $U_A(1)$ restoration~\cite{Brandt:2016daq,Tomiya:2016jwr}.

\section{Scattering and resonances within finite temperature Unitarized ChPT}

A central aspect of our analysis is the study of light thermal resonances, generated through the unitarization of scattering amplitudes at finite temperature~\cite{Dobado:2002xf}. Due to interactions with particles in the thermal environment, resonances can experience significant changes in their spectral properties. This is studied by examining the second Riemann sheet pole of the unitarized scattering amplitudes, computed within Chiral Perturbation Theory (ChPT) at finite temperature~\cite{GomezNicola:2012uc,GomezNicola:2019myi}. 

At $T=0$ and within ChPT, a two-body $a+b\to a+b$ elastic partial-wave scattering amplitude reads $t(s)=t_2(s)+t_4(s)+\cdots$, where $t_2$ is the tree-level contribution coming from the $\mathcal{O}(p^2)$ Lagrangian and $t_4$ containing one-loop diagrams from the second-order ChPT Lagrangian as well as tree-level terms from the fourth-order one proportional to the low-energy constants (LECs).
The amplitude is extended to finite temperature through the replacement $t_4(s) \rightarrow t_4(s; T)$ and then unitarized using the Inverse Amplitude
Method (IAM), resulting in
\begin{equation}
    t_{IAM}(s,T)=\dfrac{t_2(s)^2}{t_2(s)-t_4(s,T)},
\end{equation}
which satisfies exact thermal elastic unitarity \cite{GomezNicola:2023rqi}
\begin{equation}
\text{Im}\, t_{IAM}(s,T)=\left\{
\begin{aligned}
&\sigma^T_{ab}(s)\left[t_{IAM}(s,T)\right]^2, \quad s\geq (M_a+M_b)^2,\\
&\tilde{\sigma}^T_{ab}(s)\left[t_{IAM}(s,T)\right]^2, \quad 0\leq s\leq (M_a-M_b)^2,
\end{aligned}\right.
\end{equation}
where $M_a$ and $M_b$ are the masses of particle $a$ and $b$, respectively, and $\sigma_{ab}^T(s)$ and $\tilde\sigma_{ab}^T(s)$ read 
\begin{eqnarray}
    &\sigma_{ab}^T(s)=&\sigma_{ab}(s)\left[ 1+n_B\left(\frac{s+\Delta_{ab}}{2\sqrt{s}}\right)+n_B\left(\frac{s-\Delta_{ab}}{2\sqrt{s}}\right)\right],
    \label{phase1}
    \\
    &\tilde\sigma_{ab}^T(s)=&\sigma_{ab}(s)\left[ n_B\left(\frac{\Delta_{ab}-s}{2\sqrt{s}}\right)-n_B\left(\frac{s+\Delta_{ab}}{2\sqrt{s}}\right)\right],
\label{phase2}
\end{eqnarray}
with $\Delta_{ab}=M_a^2-M_b^2$ and $n_B$ the Bose-Einstein distribution function.

As in the $T = 0$ case, one has the standard unitary cut starting from the two-particle scattering threshold. At finite temperature and for $M_a\neq M_b$, a new cut appears, the so-called Landau cut. These cuts can be explained by the scattering of particles in the thermal bath. In other words, the statistical Bose-Einstein weights appearing in the thermal phase space factors \eqref{phase1}-\eqref{phase2} can be interpreted considering the stimulated production processes and the absorption by the thermal bath.

\begin{figure}[h]
    \centering
    \includegraphics[width=7.5cm]{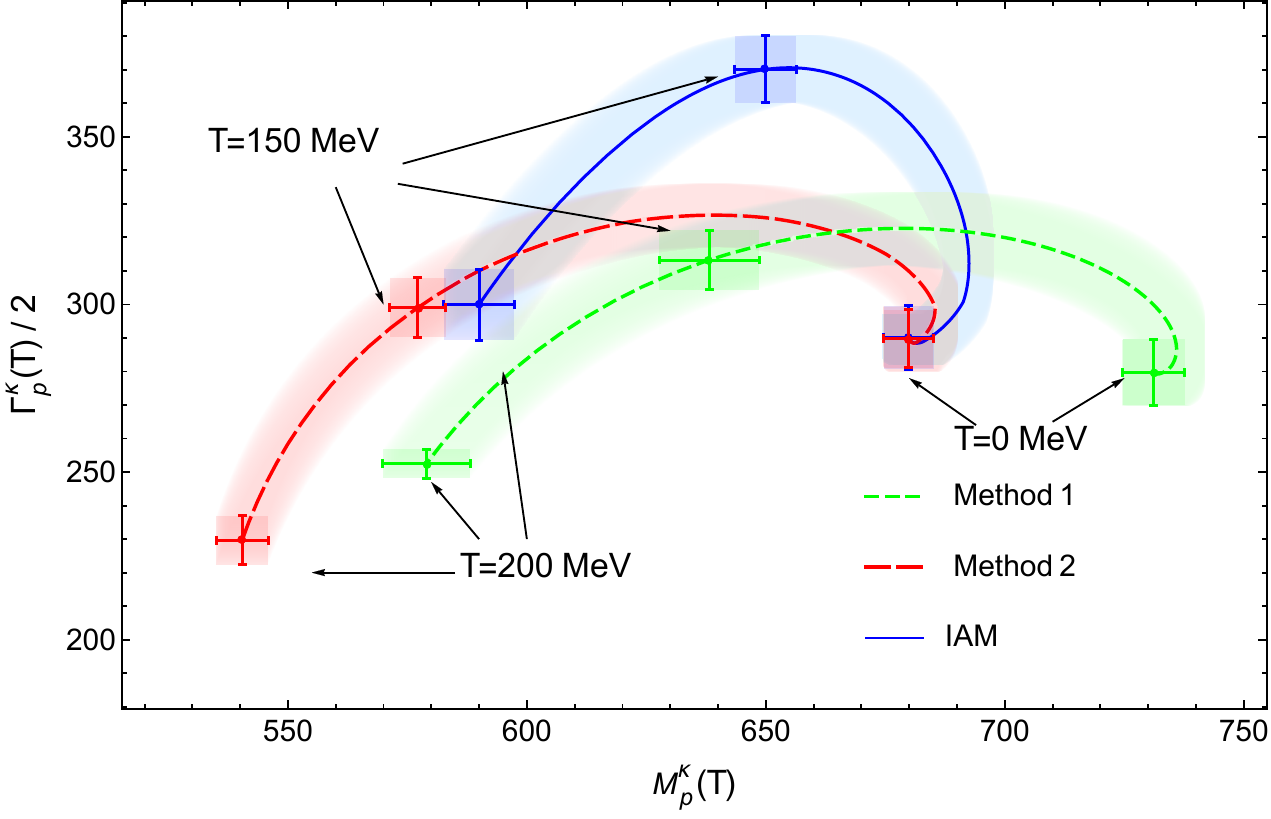}
    \includegraphics[width=7.5cm]{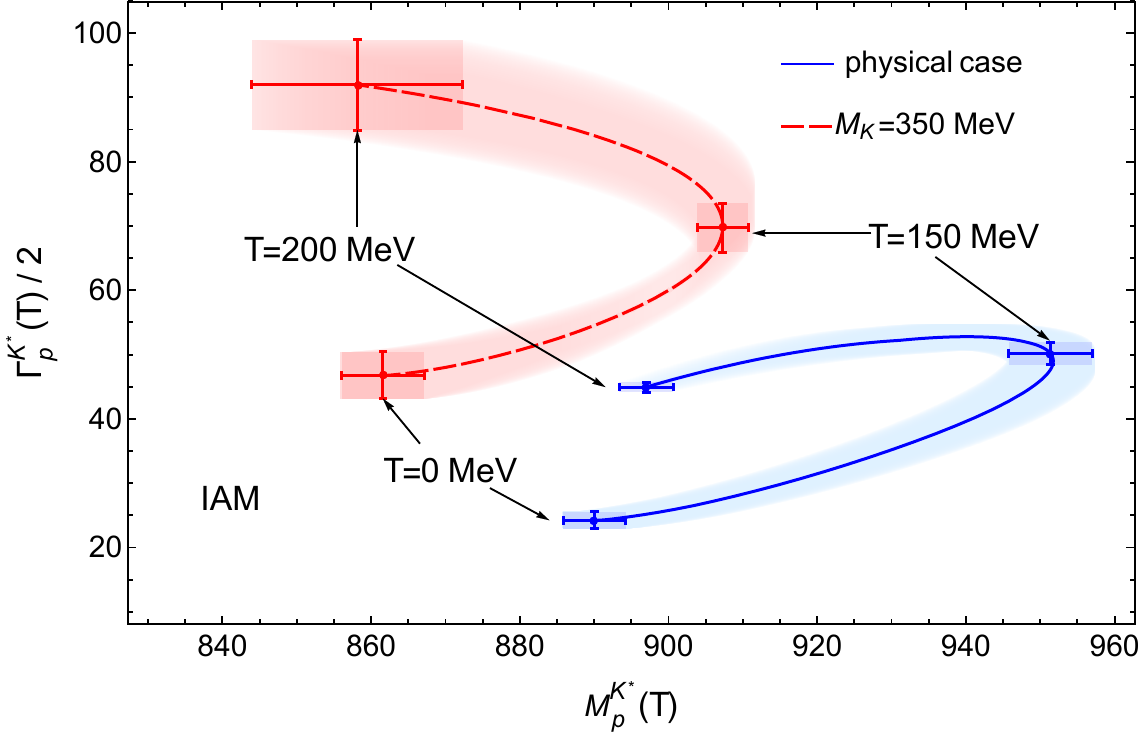}
    \caption{ Thermal dependence of the $K_0^*(700)$ (left) and $K^*(892)$ (right) poles. The pole is parameterized as customary as $\sqrt{s_p(T)} =M_p(T)-i\Gamma_p(T)/2$. The error bands correspond to the uncertainties in the LECs.}
    \label{Kmasswidth}
\end{figure}

In~\cite{GomezNicola:2023rqi}, we have computed the $\pi K$ elastic scattering amplitude and studied the thermal trajectories of the $K_0^*(700)$ and $K^*(892)$ resonances. In Figure~\ref{Kmasswidth}, we can see that the $K^*_0 (700)$ mass shows a decreasing trend at larger temperatures while its width increases at low temperatures but decreases at larger $T$ values. The $K^*(892)$ pole has a softer temperature dependence, consistent with a smoother medium dependence than for the $\rho(770)$. 

\vspace{-0.5cm}
\section{Scalar susceptibilities and light thermal resonances}

\begin{figure}[h!]
    \centering
    \includegraphics[width=8cm]{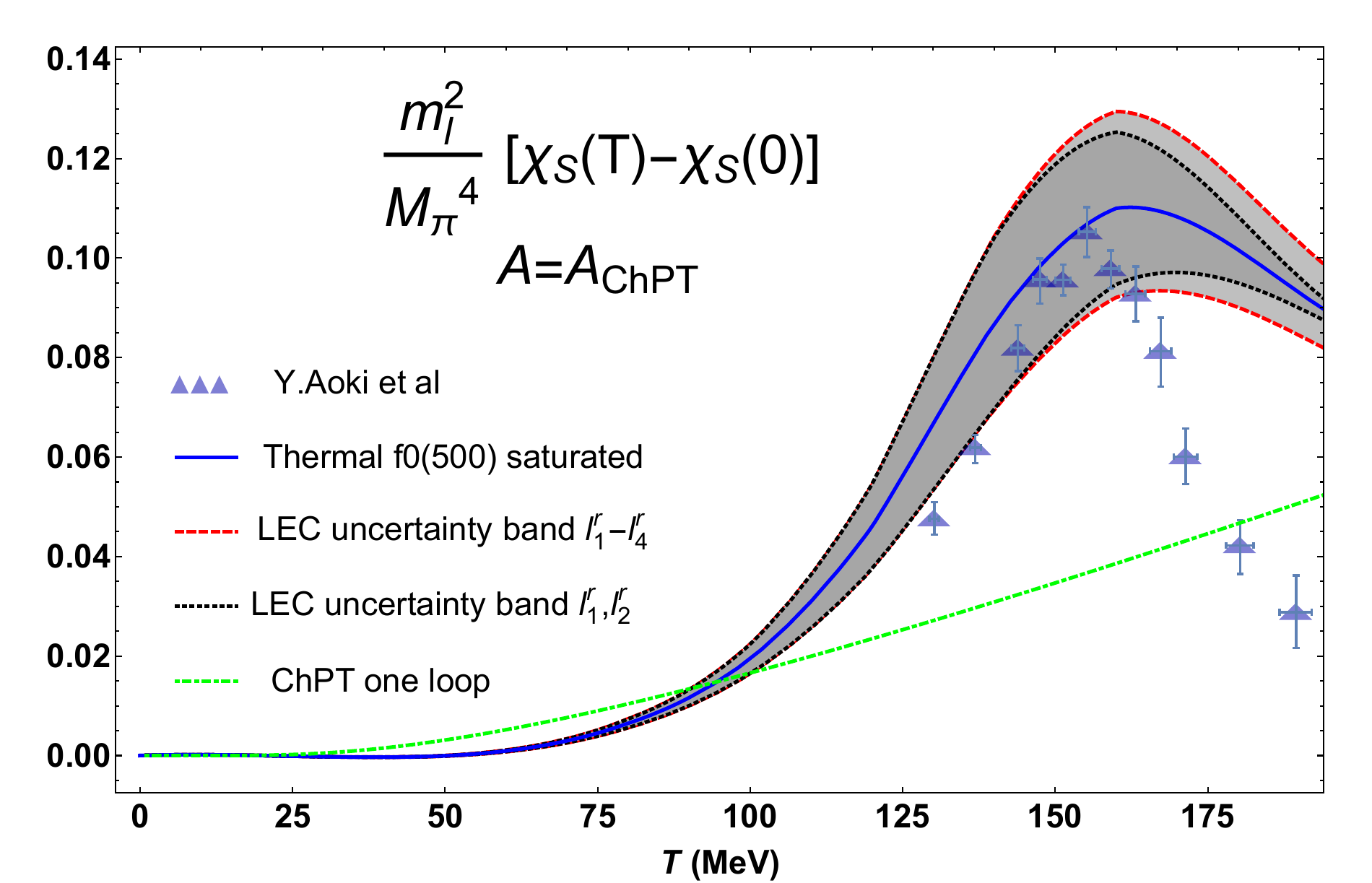}
    \vspace{-0.2cm}
    \caption{Saturated susceptibility including the LEC uncertainties given in \cite{Hanhart:2008mx}. Lattice points are from \cite{Aoki:2009sc}.}
    \label{unitsuslec}
\end{figure}

Another important outcome is the analysis of the scalar susceptibility $\chi_S$. As described in~\cite{GomezNicola:2013pgq,Ferreres-Sole:2018djq}, it can be saturated by the lightest thermal resonance $f_0(500)$
\begin{equation}
    \chi_S^U(T)=\chi_S^{ChPT}(0)\frac{M_S^2(0)}{M_S^2(T)},
\end{equation}
where $M_S^2(T)=\text{Re}\,s_p(T)$ is its thermal mass. Using the $f_0(500)$ thermal pole from the unitarized $\pi\pi$ scattering amplitude obtained at finite temperature~\cite{Dobado:2002xf}, this approach provides a robust description of the scalar susceptibility near the chiral transition. As shown in Figure~\ref{unitsuslec}, the saturated susceptibility reproduces the crossover maximum and is compatible with lattice results below and near the critical temperature, 
which highlights the close relationship between resonance dynamics and the behavior of QCD near the critical temperature. 

As an alternative sector to analyze chiral and $U_A(1)$ restoration, we have studied the isospin $1/2$ channel. The quark bilinears of the kaon and $\kappa$ mesons are related by both a chiral $O(4)$ and a $U_A(1)$ transformations, and hence, $O(4)\times U_A(1)$ restoration can be studied from $K-\kappa$ degeneration using the $I=1/2$ Ward Identities (WIs)~\cite{GomezNicola:2016ssy,GomezNicola:2020qxo}
\begin{equation}
    \chi_P^{K}(T)=\dfrac{\left[\langle\bar{q}q\rangle_l+2\langle\bar{s}s\rangle\right]}{m_s-m_l},\quad \chi_S^{\kappa}(T)=-\dfrac{\left[\langle\bar{q}q\rangle_l-2\langle\bar{s}s\rangle\right]}{m_s+m_l}.
\end{equation}
Taking the difference between the two WIs, it follows that $K-\kappa$ degeneration is controlled by the subtracted quark condensate, i.e., $\chi_P^K-\chi_S^{\kappa}\propto \Delta_{l,s}$, available in the lattice \cite{Bazavov:2011nk}. In the $m_s\gg m_l$ two-flavor limit, $\Delta_{l,s}$ is proportional to the light-quark condensate $\langle\bar{q}q\rangle_l$, implying $K-\kappa$ degeneration at the critical temperature in the exact chiral limit. In the physical case, $\Delta_{l,s}$ shows an asymptotic vanishing behavior only at larger temperatures, so the effective $O(4)\times U_A(1)$ restoration occurs only above $T_c$. Hence, $m_l/m_s$ is a relevant parameter regarding the behavior of the $K-\kappa$ susceptibility difference, and this channel helps to understand the difference between the $N_f=2+1$ and $N_f=2$ lattice results.

\begin{figure}[h!]
    \centering
    \includegraphics[width=7.5cm]{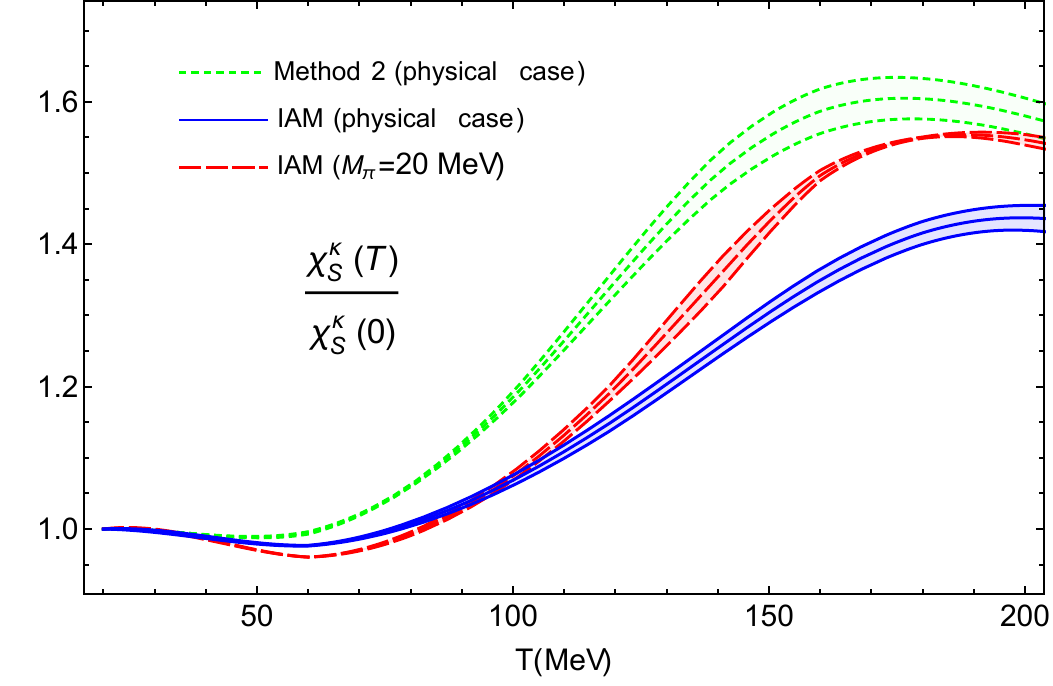}
    \includegraphics[width=7.5cm]{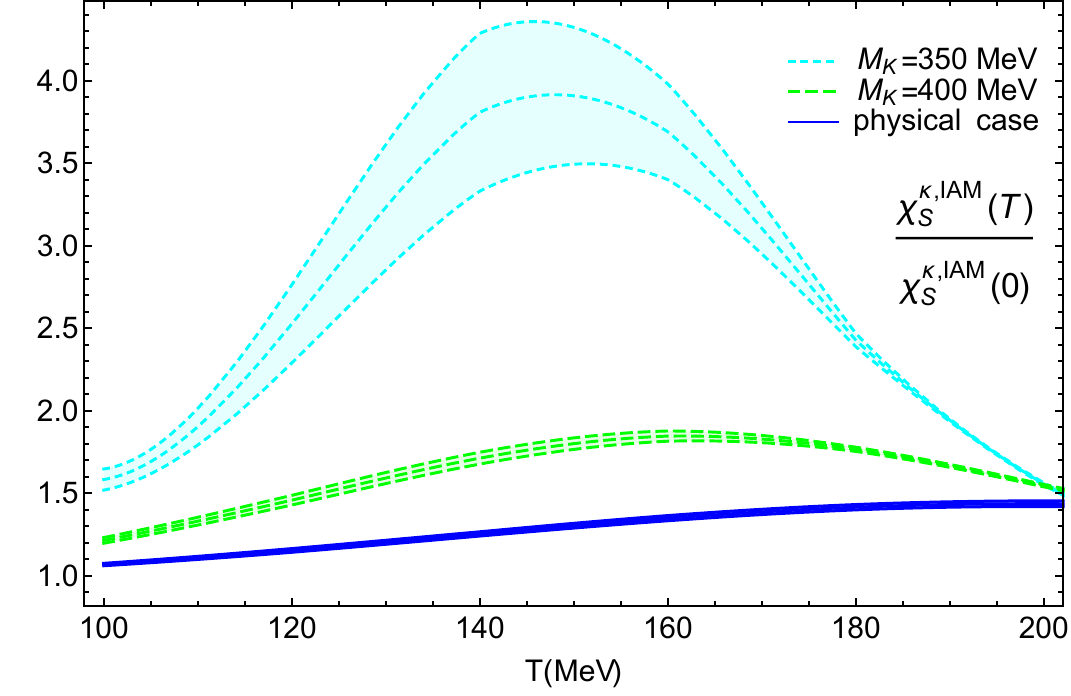}
    \vspace{-0.5cm}
    \caption{Temperature dependence of the unitarized $\kappa$ susceptibility for physical masses and the light chiral limit (left) and for $M_K$ closer to the $SU(3)$ limit (right).}
    \label{suskappa}
\end{figure}

In the $I=1/2$ channel, it is expected that the contribution of the $K_0^*(700)$ to $\chi_S^{\kappa}$ will be the most dominant because it is the lightest scalar state in this channel. Therefore, analogous to how the scalar susceptibility $\chi_S$ was saturated with the $f_0(500)$, we construct a unitarized susceptibility $\chi_S^{\kappa}$ by saturating it with the thermal $K_0^* (700)$ pole. For the unitarized $\chi_S^{\kappa}$, we get the same behavior as for the reconstructed susceptibility from lattice results using the light- and strange-quark condensates. In Figure~\ref{suskappa}, we can see the peak of the $\chi_S^{\kappa}$ scalar susceptibility at a temperature above the chiral critical temperature. We can also note that near the light-chiral limit ($M_\pi=20$ MeV) the growth of the curve below the peak is more pronounced and compatible with a flattening above the peak, which seems to indicate $K-\kappa$ degeneration at lower temperatures. Close to the $SU(3)$ limit ($M_K=350$ MeV), the peak displaces towards $T_c$ and growths, consistent with the degeneration of $\chi_S^{\kappa}$ with $\chi_S$. 

\section{Pion scattering and critical temperature at nonzero chiral imbalance}

Finally, we have explored the evolution of chiral restoration for nonzero chiral imbalance chemical potential $\mu_5$. For this purpose, we have constructed the most general effective Lagrangian for a nonzero axial chemical potential $\mu_5$ \cite{Espriu:2020dge} through the so-called external source method. This Lagrangian includes different types of terms coming from the covariant derivatives and explicit source terms. While the $\mu_5$ dependence of the second-order Lagrangian is a constant term, the fourth-order  chiral Lagrangian is modified in the following way
\begin{equation}
    \mathcal{L}_{4}(\mu_{5}) = \mathcal{L}_{4}(\mu_{5}=0) + \kappa_{1}\mu_{5}^{2}\text{tr}(\partial^{\mu}U^{\dagger}\partial_{\mu}U) + 
    \kappa_{2}\mu_{5}^{2}\text{tr}(\partial_{0}U^{\dagger}\partial^{0}U) + \kappa_{3}\mu_{5}^{2}\text{tr}(\chi^{\dagger}U+U^{\dagger}\chi)+ \kappa_{4}\mu_{5}^{4}.
    \label{lagp4mu5}
\end{equation}

From the energy density obtained within the above equation \eqref{lagp4mu5},  we derive the quark condensate. This quantity can be used to extract some information about the $\kappa_i$ LECs involved from the critical temperature determined as the value for which the quark condensate vanishes. At NNLO, the light quark condensate normalized to its zero temperature value depends only on the combinations $\kappa_a=2\kappa_1-\kappa_2$ and $\kappa_b=\kappa_1+\kappa_2-\kappa_3$. On the other hand, the chiral critical temperature can be defined as the temperature at which the scalar susceptibility has a maximum, so we can calculate this one from the saturated approach mentioned in the previous section. To obtain the $\mu_5$ dependence of the saturated susceptibility, we have calculated the $\mu_5$ corrections to the unitarized $\pi\pi$ elastic scattering amplitude within the ChPT framework at finite temperature \cite{GomezNicola:2023ghi}. The $\mu_5$ dependence of the $IJ=00$ perturbative ChPT partial wave depends on two different $\kappa_i$ combinations: $\kappa_1'=6\kappa_1+5\kappa_2$ and $\kappa_2'= -8\kappa_1-4\kappa_2+5\kappa_3$.

\begin{figure}[h!]
    \centering
    \includegraphics[width=7cm]{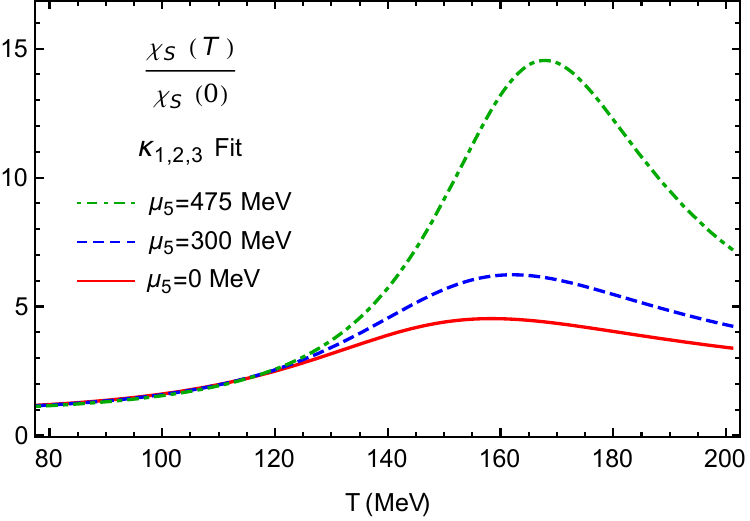}
    \includegraphics[width=8cm]{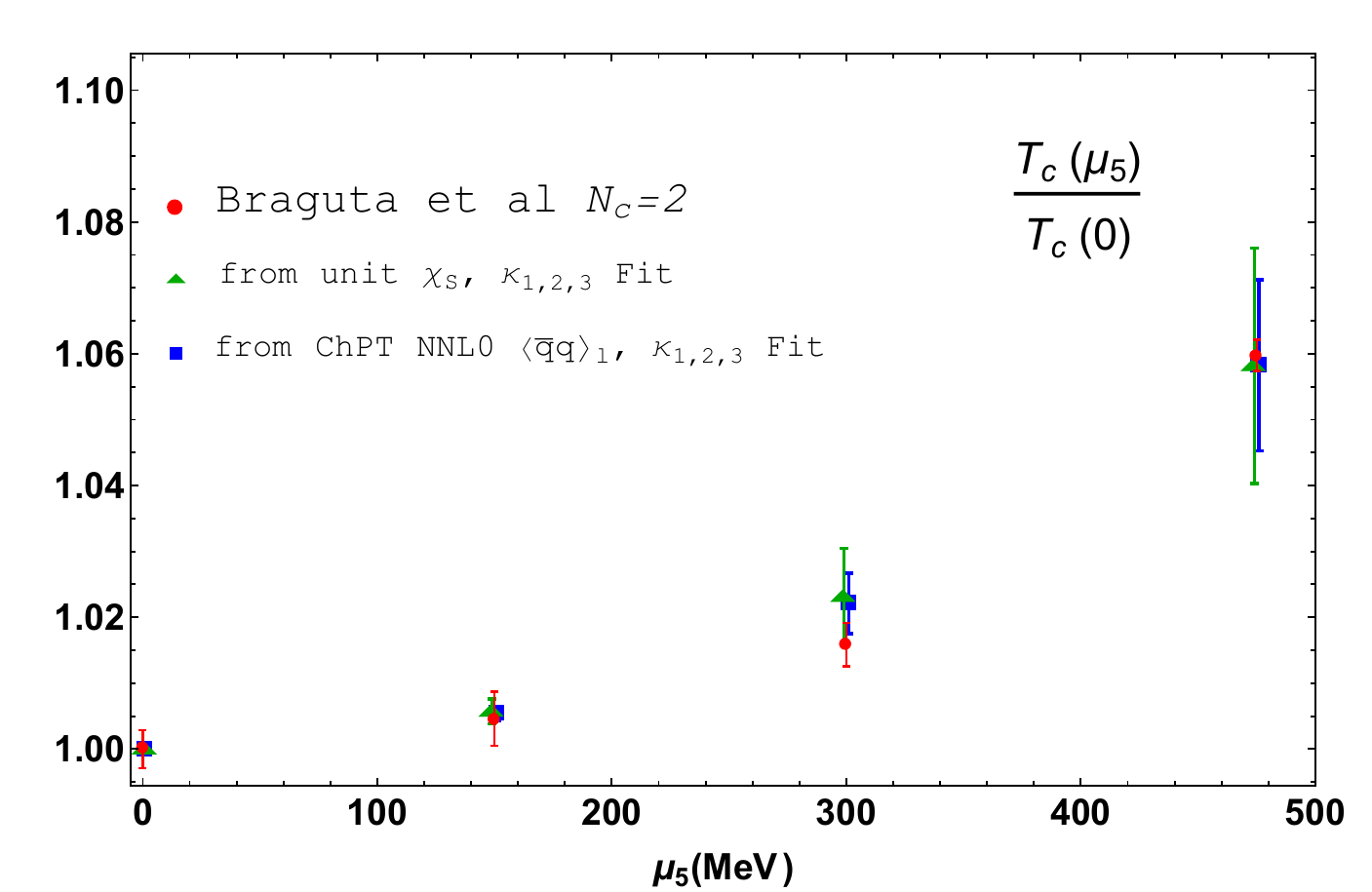}
    \vspace{-0.7cm}
    \caption{Left: Right: Combined fit of $Tc(µ5)/Tc(0)$ determined from the NNLO ChPT quark condensate, the saturated susceptibility, and the topological susceptibility. We also show the lattice results from \cite{Braguta:2015zta}.}
    \label{susmu5}
\end{figure}

We have carried out a combined fit with $\kappa_{1,2,3}$ as fit parameters where the fitted observables are the critical temperature with both methods mentioned, namely, the ChPT condensate and the scalar susceptibility, and the topological susceptibility analyzed in \cite{Espriu:2020dge}. The result of the fit is $\kappa_1=9.4^{+1.1}_{-1.3}$, $\kappa_2=-4.5^{+1.5}_{-1.4}$ and $\kappa_3=3.6^{+9.1}_{-8.7}$ where the uncertainties corresponds to the $95\%$ confidence level. Figure \ref{susmu5} shows the saturated susceptibility and the critical temperature plotted with the $\kappa_i$ set corresponding to our fit. As we can see, lattice points are compatible with both $T_c$ determinations, confirming the growing behavior of the critical temperature when chiral imbalance is increased.

\section{Conclusions}

We have demonstrated the importance of the $f_0(500)$ and $\kappa$ thermal resonances in chiral and $U_A(1)$ symmetry restoration. Our results show that the mass and the width of these resonances undergo significant shifts with increasing temperature,  consistent with the expected behavior near the chiral transition. In addition, the $\chi_S$ and $\chi_S^{\kappa}$ saturated susceptibilities reveal a behavior consistent with the lattice results. Finally, we have analyzed chiral restoration in the presence of chiral imbalance and found that the chiral critical temperature increases with $\mu_5$, corroborating lattice studies.

\section*{Acknowledgments}

Work supported by the Ministerio de Ciencia e Innovación, research contract PID2022-136510NB-C31, the European Union Horizon 2020 program under grant agreement No 824093, and the Ramón y Cajal program (RYC2019-027605-I) of the Spanish MINECO.

\end{document}